\documentclass[conference]{IEEEtran}
\IEEEoverridecommandlockouts
\usepackage{cite}
\usepackage{amsmath,amssymb,amsfonts}
\usepackage{algorithmic}
\usepackage{graphicx}
\usepackage{textcomp}
\usepackage{xcolor}
\usepackage{booktabs}
\usepackage{array}
\usepackage{hyperref}

\usepackage{capt-of}
\usepackage[table]{xcolor}
\usepackage{makecell}

\hypersetup{
    hidelinks
}
\def\BibTeX{{\rm B\kern-.05em{\sc i\kern-.025em b}\kern-.08em
    T\kern-.1667em\lower.7ex\hbox{E}\kern-.125emX}}
\begin{document}

\title{
The Help Ladder: Skill-Adaptive Peer Scaffolding for Real-Time Collaborative Programming
\\
\thanks{This work was supported by the Center for Human-Computer Interaction at Virginia Tech (CHCI).}
}

\author{
\IEEEauthorblockN{
Panayu Keelawat\textsuperscript{\dag},
Sriram Narlapati\textsuperscript{\dag},
Sangwook Lee\textsuperscript{\dag},
Darshan Nere\textsuperscript{\dag},\\
Griffin Ogura\textsuperscript{\dag},
Xinran Adeline Li\textsuperscript{\ddag},
Sang Won Lee\textsuperscript{\dag},
Yan Chen\textsuperscript{\dag}
}

\vspace{1ex}

\IEEEauthorblockA{
\textsuperscript{\dag}\textit{Virginia Tech}, USA\\
\{panayu, nsriram, sangwooklee, darshannere, gogura, sangwonlee, ych\}@vt.edu
}

\IEEEauthorblockA{
\textsuperscript{\ddag}\textit{Johns Hopkins University}, USA\\
xli436@jhu.edu
}
}

\newcommand{\sys}[1]{\textsc{Canary}}

\maketitle

\begin{abstract}
Collaborative programming is a widely adopted classroom activity to encourage peer scaffolding, yet real-time collaboration often breaks down into parallel individual work with minimal interaction. Our formative studies reveal that even when students want to collaborate, they are held back by the effort required to understand a teammate’s entire problem at once. We present Canary, a system that supports peer scaffolding by breaking down programming obstacles into smaller steps tailored to a student's skill level. Canary alerts potential helpers to specific places where they can start, using AI to turn complex problems into a step-by-step ladder that starts with easy fixes before moving toward harder logic. By providing this gradual ramp-up, Canary enables students to make quick contributions and progressively work toward solving their peers’ problems. Our evaluation shows that this staged approach makes helping feel less overwhelming, leading to more frequent and effective collaboration among students.
\end{abstract}


\section{Introduction}

Collaborative programming (CP) has seen increasingly wide adoption in computer science classrooms, as it has been shown to enhance students' problem-solving abilities, engagement, and knowledge retention \cite{collabEduEvidence}. In these settings, students are typically organized into small groups of three to four members to work on a shared codebase, such as object-oriented programming (OOP) tasks \cite{rcpForEdu}, which can be decomposed into separate methods assigned to different team members while having interdependencies among function calls within the system. This shared workspace is intended to foster an environment where students collaborate and learn with peers, creating opportunities for peer scaffolding in which they teach and support one another. Despite the rise of generative AI, solving problems collaboratively continues to benefit both helpers, who reinforce their understanding by explaining concepts, and helpees, who receive targeted assistance \cite{collabLearningInProg, crouch2001peer}. Real-time collaborative programming (RCP) tools, such as VS Code Live Share, have demonstrated their potential to improve programming performance among undergraduate students \cite{rcpInUndergrad}. However, despite the availability of these shared environments, recent studies highlight a persistent ``silo effect'': many groups exhibit minimal interaction rather than engaging in meaningful collaboration, undermining the intended learning benefits of these activities \cite{collabCS101, selfSelectedVsRandomGroups}.

\begin{figure}
    \centering
    \includegraphics[width=1\linewidth]{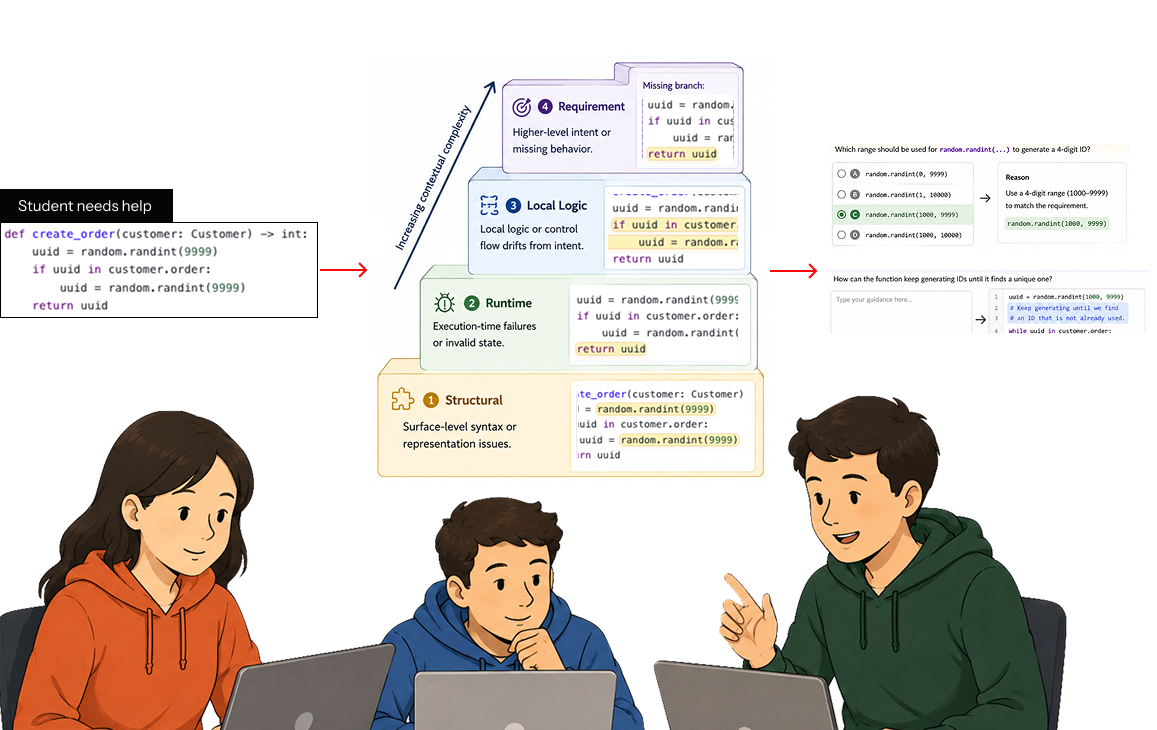}
    \caption{We present \sys{}, a collaborative programming environment that supports peer scaffolding through progressive contextual guidance. When a student encounters difficulty during real-time collaborative programming, \sys{} decomposes the obstacle into a layered \textit{Help Ladder} consisting of structural, runtime, local logic, and requirement-level issues. This allows helpers to enter through lightweight, low-context contributions before progressively engaging with deeper reasoning about the teammate’s implementation.}
    \label{fig:teaser}
\end{figure}

To better understand the challenges faced by CP teams, we conducted a series of formative studies grounded in the setting of a programming course at our institution, where in-class team activities involve groups of three students working on shared OOP problems in Python. First, we recruited 30 students (ten teams) and examined their collaboration behaviors when using traditional RCP tools through observations and post-task interviews. We found that teams exhibited minimal interaction, as students who encountered difficulties rarely reached out to their peers for help. Participants reported hesitating to seek assistance because they did not want to be perceived as not contributing, and instead preferred to continue trying independently. We then conducted a preliminary study with 12 students (four teams) using an early prototype of our system, which proactively recommended help sessions by reaching out to available teammates (e.g., those who had just completed a task) on behalf of students who were stuck. Despite this intervention, helpers often ignored these recommendations and continued working on their own tasks. Follow-up interviews revealed that while participants were all willing to help, they found it difficult to quickly understand others’ code and its relation to test cases, especially for unfamiliar tasks. Additionally, participants were reluctant to interrupt teammates who appeared to be actively working, due to concerns about disrupting their workflow. While prior work has explored the use of LLMs for algorithm learning scaffolding~\cite{dbox}, there remains a gap in layering help requests to support helpers in quickly understanding the problem context and enabling effective peer scaffolding within real-time collaborative programming teams.

We therefore propose \sys{}, a shared IDE that supports peer scaffolding through LLM-powered layered help requests. \sys{} uses a step-by-step approach to support complex problem solving \cite{dbox, InstructAR}. Rather than presenting helpers with the full complexity of a teammate's issue at once, \sys{} decomposes programming obstacles into a sequence of progressively more complex steps. It alerts potential helpers to specific, localized starting points and provides a structured progression from easy, surface-level fixes to deeper logical errors. When multiple issues exist at the same level, \sys{} prioritizes those most relevant to console output as well as those aligned with the helper’s expertise, which it models by tracking users’ accumulated concepts through real-time problem solving. This layered approach lowers the barrier to entry for helping, enabling students to make quick initial contributions to get their foot in the door and progressively engage with more complex aspects of the problem. By bridging the helpee’s work with the helper’s context, \sys{} aims to reduce the cognitive cost of understanding a teammate’s code. Helpers can use the provided context to formulate responses by manually typing in the system or to support verbal discussion with the helpee.

We evaluated \sys{} through a within-subject study comparing it against a baseline collaborative IDE for RCP without structured help requests, involving 12 CS students (four teams). Our results show that teams using \sys{} engaged in a greater quantity of peer scaffolding (i.e., more collaboration sessions) and successfully unblocked more issues. We also found that enabling more effective peer scaffolding led to a higher number of tasks completed overall compared to the baseline. Participants additionally reported lower collaboration workload.

To summarize, our work makes the following contributions:
\begin{itemize}
    \item \sys{}, a collaborative IDE for RCP that supports peer scaffolding through LLM-powered layered help requests.
    \item A progressive interface design for structuring help requests, enabling helpers to more easily engage with a teammate’s code problem.
    \item User study results demonstrating that \sys{} fosters more active and effective collaboration in programming teams.
\end{itemize}

\section{Related Work}

\subsection{Collaborative Programming and Peer Learning}

CP is widely adopted in computer science education to improve engagement and performance~\cite{collabEduEvidence, csclInProgEdu}. Active learning approaches such as peer instruction and guided inquiry show that students learn effectively by explaining concepts and solving problems collaboratively~\cite{crouch2001peer, POGIL}. RCP tools further support this by enabling synchronous interaction and shared code manipulation, improving problem-solving performance and programming outcomes~\cite{rcpInUndergrad}.

Despite these benefits, collaboration often falls short in practice. Students frequently work in parallel with limited interaction, influenced by group dynamics and composition~\cite{collabCS101, selfSelectedVsRandomGroups}. Social factors further inhibit collaboration: students may hesitate to seek help due to fear of judgment or concerns about competence~\cite{helpSeekingInUni}, and may exhibit unproductive help-seeking behaviors~\cite{unproductiveHelpSeeking}.

Prior work has explored mechanisms to improve collaboration through awareness and coordination support. For example, personal-space designs increase engagement by promoting task ownership~\cite{devPersonalSpace}, while gaze-awareness systems reduce the effort required to reference code during collaboration~\cite{pairProgGazeAwareness}. Other systems surface collaboration signals to instructors to support targeted intervention in student-group activities~\cite{VizGroup, Dynamite}. However, these approaches primarily focus on coordination and awareness rather than reducing the cognitive effort required for collaborators to quickly understand and engage with a teammate’s programming problem.

\subsection{Team Awareness and Coordination Systems}

A central challenge in collaborative systems is maintaining awareness of teammates' activities and progress~\cite{workspaceAwareness}. Tools such as FASTDash visualize team activity to improve awareness of code changes~\cite{FASTDash}, while studies of RCP platforms highlight the need for features such as focus-follow and shared execution to support collaboration~\cite{VSCodeLiveShare}.

Beyond awareness, systems have embedded coordination support into workflows. TaskBot reduces context switching by integrating task management into chat~\cite{chatbotCHI}, while CHOIR organizes team knowledge using LLMs~\cite{CHOIR}. Orchestration Scripts provide situated guidance by encoding work practices~\cite{OrchestrationScripts}, and other systems leverage ambient cues or shared displays to enhance coordination~\cite{cyclingSharing, facialCuesCollab, Novecs, LADICA}.

While these systems improve visibility and coordination, they do not address the cognitive effort required to understand a teammate's problem. In collaborative programming, awareness of a problem is insufficient if helpers cannot quickly interpret and act on it. Our work addresses this gap by embedding localized and progressively structured help requests directly into the IDE.

\subsection{AI-Powered Programming Support and Scaffolding}

LLM-based programming assistants support code generation, explanation, and debugging~\cite{copilotNoviceStudy, CodeAid}. Systems such as Ivie improve comprehension through in-situ explanations~\cite{Ivie}, while DBox scaffolds problem decomposition through structured guidance~\cite{dbox}. In collaborative contexts, systems like CoPrompt support shared prompt interaction~\cite{CoPrompt}, and Codeon distributes subtasks among contributors to support collaborative problem solving~\cite{Codeon}.

Research on human-AI collaboration highlights trade-offs between efficiency and human agency~\cite{humanVsAiPairProg, humanToAgentPair, AIPairProgramming}. Prior work emphasizes the importance of maintaining cognitive engagement, avoiding full solutions, and mitigating disruption from proactive assistance~\cite{CodeAid, Codellaborator, genAIAndCT}.

Unlike prior systems that focus on assisting individual users or generating solutions, \sys{} supports the helper in peer collaboration. By generating layered help requests grounded in code context and user expertise, \sys{} shifts AI assistance from direct problem solving toward enabling effective human-to-human scaffolding.

\section{Needfinding Study}

To better understand the challenges of real-time collaboration, we conducted an in-lab study modeled after an existing in-class activity in a CS course at our institution. The goal was to examine how students coordinate in ad-hoc teams and to uncover the factors shaping their collaboration behaviors.

\begin{figure}
    \centering
    \includegraphics[width=1\linewidth]{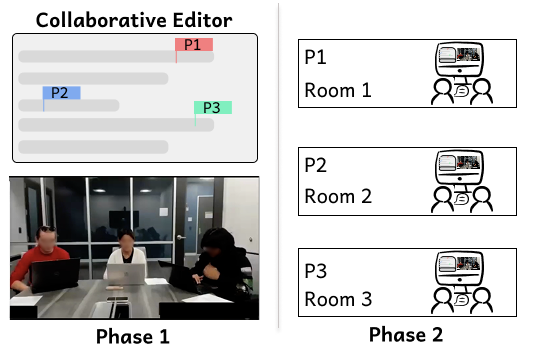}
    \caption{Needfinding procedure consisting of Phase 1 (collaborative programming) and Phase 2 (post-hoc interviews). Example from one group with participants P1–P3.}
    \label{fig:needfindingStudyFig}
\end{figure}

We recruited 30 students with prior Python programming experience from our institution (P1--P30) and randomly assigned them into 10 teams of three members each (e.g., Team 1: P1--P3, Team 2: P4--P6, etc.). The study consisted of two phases (Fig.~\ref{fig:needfindingStudyFig}).

In Phase 1, each team was given a programming challenge to implement a number-guessing game consisting of three required functions. Teams collaborated in VS Code using the Live Share feature, which displayed each member’s cursor in real time. Participants were not allowed to use AI coding assistants. We screen-recorded the coding sessions and video-recorded the participants. This phase lasted 20 minutes.

In Phase 2, each participant was interviewed individually in a separate room. The interviews focused on clarifying their internal thought processes during collaboration, particularly during moments of silence or hesitation observed in Phase 1. Each interview lasted approximately 15 minutes.

We analyzed screen recordings, interview transcripts, and code artifacts produced by each team. 

Only 3 out of 10 teams produced fully working solutions that passed all predefined test cases. All teams adopted a divide-and-conquer strategy early on, assigning each member a separate function. Following this, participants primarily focused on their own tasks. Across teams, we observed a consistent pattern: when participants encountered difficulties, they tended to persist in working individually rather than seeking help from teammates. Interviews revealed that students were reluctant to seek help even when stuck, due to being immersed in their own workflow, social pressure, and concerns about interrupting others. For example, P15 noted, ``\textit{I was like nervous. I was scared [of] the [remaining] time. ...Maybe [the team could] handle it... faster,}'' reflecting hesitation to engage others out of concern that doing so might disrupt their teammates' progress. Participants also reported limited awareness of their teammates’ progress. As P1 described, ``\textit{[Everyone] did work on separate parts, but we didn’t really know what each other were doing.}'' This lack of visibility made it difficult to recognize when help was needed.

Overall, these findings suggest that students were hesitant to seek help even when they were stuck. Uncertainty about when it was appropriate to intervene, combined with a desire to avoid disrupting teammates, led many to continue working in isolation, delaying or missing opportunities for peer scaffolding.

\section{Preliminary Study}

\begin{figure}
    \centering
    \includegraphics[width=1\linewidth]{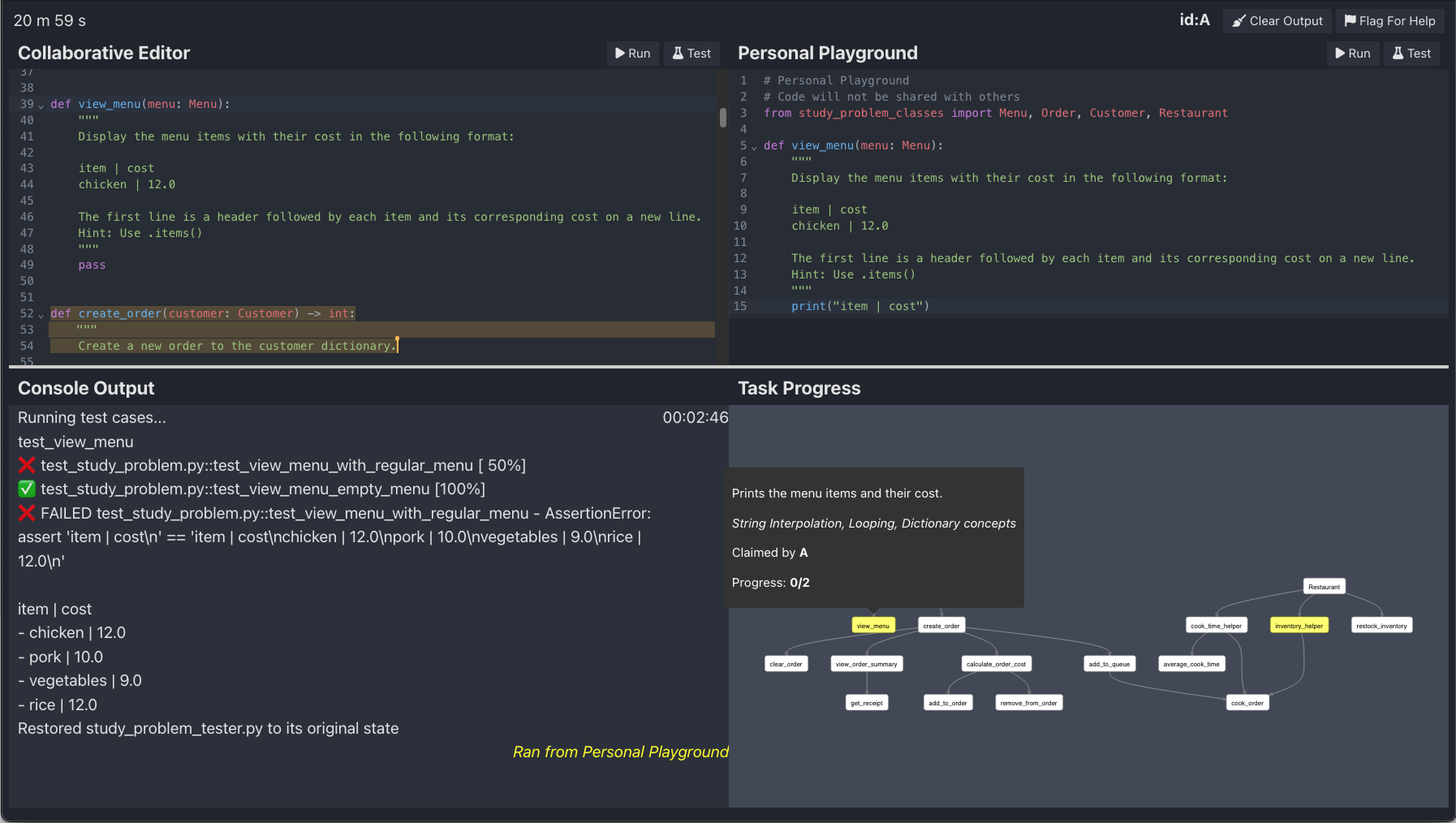}
    \caption{User interface of the early version of the system used in the preliminary study.}
    \label{fig:oldUI}
\end{figure}

Based on insights from the needfinding study, we developed an early prototype of our system (Fig.~\ref{fig:oldUI}) and conducted a preliminary study with 12 CS students (four teams of three, randomly assigned) following the same study protocol. The prototype provided a visual task breakdown of the assignment as a dependency graph and displayed which team member was working on each function. It also tracked individual progress and identified teammates who appeared to need help, inferred from prolonged inactivity or exceeding an estimated completion time. When a user completed their task, the system pushed a notification suggesting them to assist a teammate before picking up a new task.

Despite these features, collaboration did not meaningfully improve, with groups averaging only 1.25 help sessions and 0.50 successfully unblocked issues per session. Potential helpers were often hesitant to initiate collaboration. Interviews revealed that although participants were all willing to help their teammates, they were frequently uncertain about whether they possessed the relevant expertise to contribute effectively and were concerned about wasting time if their assistance was not useful. In addition, because help requests were initiated by the system rather than explicitly by the helpee, helpers often assumed that the helpee was still making progress independently. As a result, they frequently chose to move on to new tasks rather than intervene. Further analysis of code artifacts at the moment when notifications were triggered showed that helpees often had multiple issues in their code at the same time, ranging from trivial problems (e.g., syntactic errors or typos in print statements) to more complex logical issues.

These findings suggest that conventional UI cues are insufficient for conveying the underlying causes of a teammate’s difficulty in real-time collaborative settings. Understanding another student’s issue often required substantial contextual comprehension under time pressure, leading helpers to default to independent work rather than engage in potentially costly coordination.

\section{Design Goals}

Motivated by our formative studies, we derive the following design goals to guide the development of \sys{}:

\begin{itemize}
    \item \textbf{DG1: Proactively initiate help opportunities.} The system should suggest help sessions by reaching out on behalf of helpees, inferring potential struggles from real-time editor activities to initiate collaboration.
    \item \textbf{DG2: Reduce the cost and disruption of peer assistance.} The system should enable helpers to make quick, low-effort contributions while minimizing interruption to both the helper’s and helpee’s ongoing workflows, supporting lightweight and time-efficient collaboration.
    \item \textbf{DG3: Preserve helper knowledge contribution.} The system should encourage helpers to actively contribute their own understanding and expertise, fostering peer scaffolding rather than supplying direct solutions for helpers to simply pass along.
    \item \textbf{DG4: Provide sufficient contextual grounding.} The system should supply helpers with relevant context to reduce cognitive load, minimize context switching, and increase confidence in engaging with a teammate’s problem.
\end{itemize}

\section{\sys{}}

\subsection{Interface Overview}

\begin{figure*}
    \centering
    \includegraphics[width=1\linewidth]{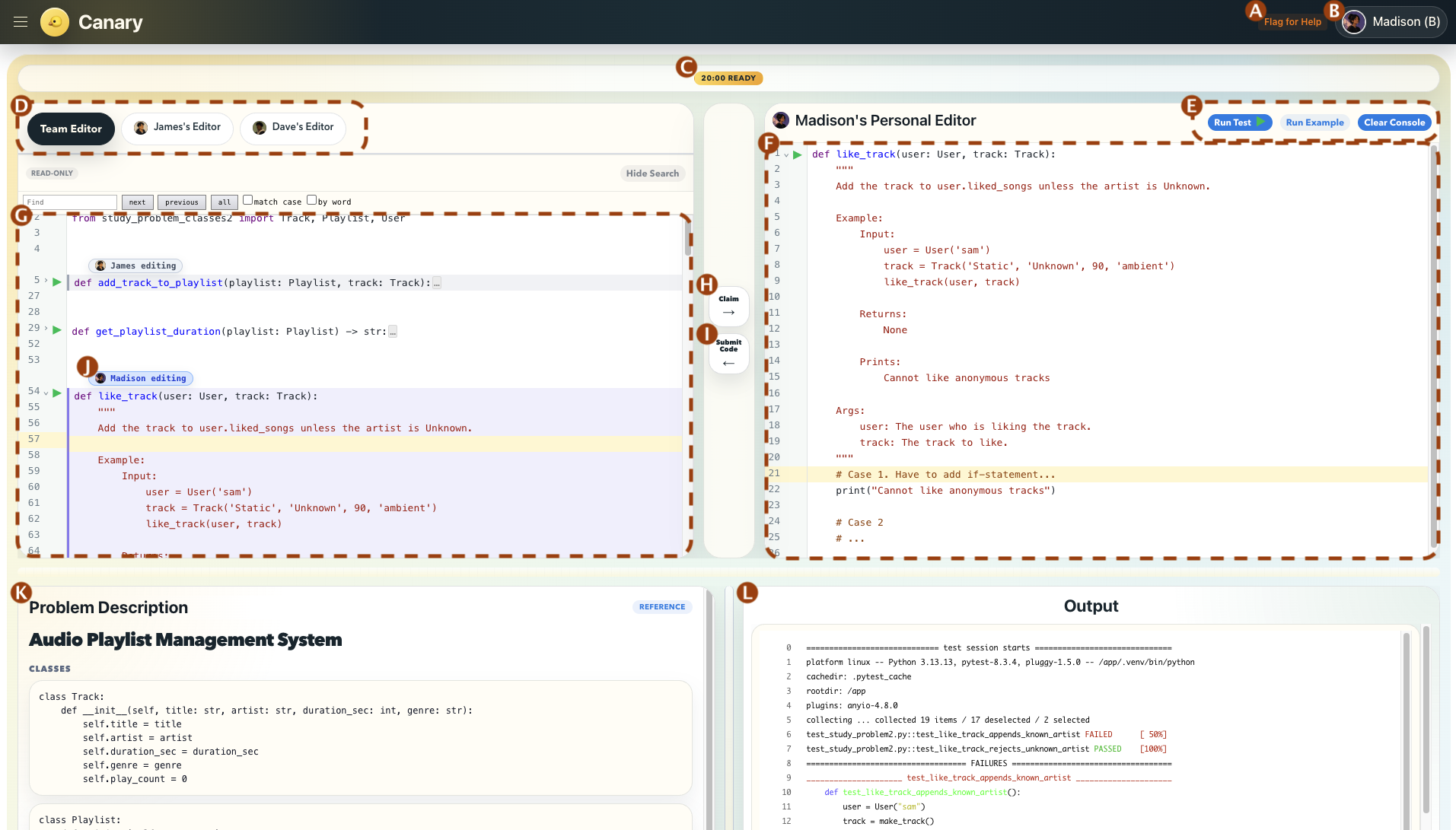}
    \caption{Interface of \sys{}. (A) \textit{Flag for Help} button for explicitly requesting assistance from peers. (B) Current user indicator displaying the user’s profile picture, name, and participant ID. (C) Shared session timer showing the remaining activity time. (D) Tabs for navigating the shared team editor codebase and viewing teammates’ personal editors in read-only mode. (E) Controls for running tests, executing custom inputs for debugging, and clearing console output. (F) Personal editor where users independently implement and debug their assigned functions. (G) Shared editor space with real-time awareness indicators of teammates’ editing activity. (H) \textit{Claim} button for claiming ownership of a function from the shared workspace and copying it into the personal editor. (I) \textit{Submit Code} button for pushing completed implementations back to the shared team editor. (J) Function-level ownership and editing awareness indicators embedded within the shared editor. (K) Problem description panel containing assignment requirements and reference materials. (L) Personal output console displaying execution logs and test feedback for the current user.}
    \label{fig:overallInterface}
\end{figure*}

The \sys{} workspace is designed to support RCP through a structured, multi-panel interface. As shown in Fig.~\ref{fig:overallInterface}, the workspace consists of four primary panels. The top row contains two editors: a shared team editor on the left and a personal editor for the user on the right. The bottom row includes the problem description and the output logs.

The team editor provides a read-only view of the shared codebase, where users can browse functions and observe their teammates’ progress in real time. Multiple team editors, including peers' views, are accessible through tabs. To maintain clarity of collaboration, each function in the team editor is annotated with ownership and status indicators, showing which teammate is currently working on it. This design supports awareness of team activity without introducing editing conflicts.

To work on a task, users select a function from the team editor and click the \textit{Claim} button located between the two editors. This action copies the selected function into the user’s personal editor, where they can freely modify the code. The personal editor serves as an isolated workspace, allowing users to focus on their assigned task without interfering with others' work.

Users can test their implementations by clicking the \textit{Run Test} button or the green execution icon next to function names, which runs the code against predefined test cases. The results are displayed in the output logs panel below. Once satisfied with their solution, users can click \textit{Submit Code} to push their changes back to the shared team editor.

\subsection{Proactive Identification of Help Opportunities}

\sys{} continuously monitors each team member’s progress to infer when a student may be struggling and in need of assistance. When such situations are detected, the system flags the student as a potential helpee and queues them for future peer scaffolding opportunities, enabling the system to later connect them with appropriate helpers (DG1). \sys{} adapts proactive coordination features from prior work~\cite{Codellaborator} to support its design. Rather than directly triggering immediate collaboration, the system maintains a dynamic representation of who may need help and on which tasks, which is later used to surface entry points for potential helpers.

To determine whether a student is stuck, \sys{} leverages multiple signals derived from real-time programming activity. First, it performs lightweight semantic analysis of code comments to detect expressions of confusion (e.g., ``\texttt{\# How to print a float with 2 decimal points?}''), evaluated at one-minute intervals. Second, it monitors execution logs to identify repeated failures; if the same error occurs at least three times within a two-minute window, the system infers that the student may be unable to resolve the issue independently. Third, it considers time-on-task: if a student spends more than seven minutes actively working on the same function, the system flags this as a potential struggle. In addition to these automated signals, students can explicitly request help at any time using a \textit{Flag for Help} button, which marks the current task as needing assistance. If the student resolves the issue on their own, the system clears the help-needed status accordingly.

Once helpees are flagged as needing assistance, \sys{} determines appropriate moments to surface scaffolding opportunities. The system presents these opportunities to other team members at natural breakpoints in workflow, such as when a student completes a task and begins looking for a new one, or when a period of inactivity (e.g., no keyboard input for 30 seconds) is detected.

\subsection{The Help Ladder}

To address the high cognitive barrier identified in our formative studies, \sys{} implements a \textit{Help Ladder}, a layered scaffolding mechanism that structures peer assistance according to the amount of contextual understanding required from the helper (DG2). The system models helping as a progressive process of context acquisition. The design draws on scaffolding theory, which suggests that effective support should reduce task complexity and gradually adapt as learners develop understanding \cite{wood1976role, van2010scaffolding}. \sys{} also incorporates the ``foot-in-the-door'' principle~\cite{footInTheDoor} by encouraging helpers to begin with lightweight, low-commitment forms of assistance before progressively engaging in deeper reasoning and abstraction. The ladder incrementally reveals increasingly contextualized support opportunities, enabling helpers to contribute through localized insights before transitioning to more comprehensive problem-solving.

\sys{} organizes programming obstacles into four levels, each requiring increasing amounts of contextual understanding from the helper (Fig.~\ref{fig:levels}). This design is informed by prior work suggesting that different problem-solving demands require different forms of support and guidance \cite{debuggingWithAITutor}. We therefore derived these categories from the formative study code artifacts based on the amount and type of contextual information needed to understand and address each issue.

The levels include:

\begin{figure*}
    \centering
    \includegraphics[width=1\linewidth]{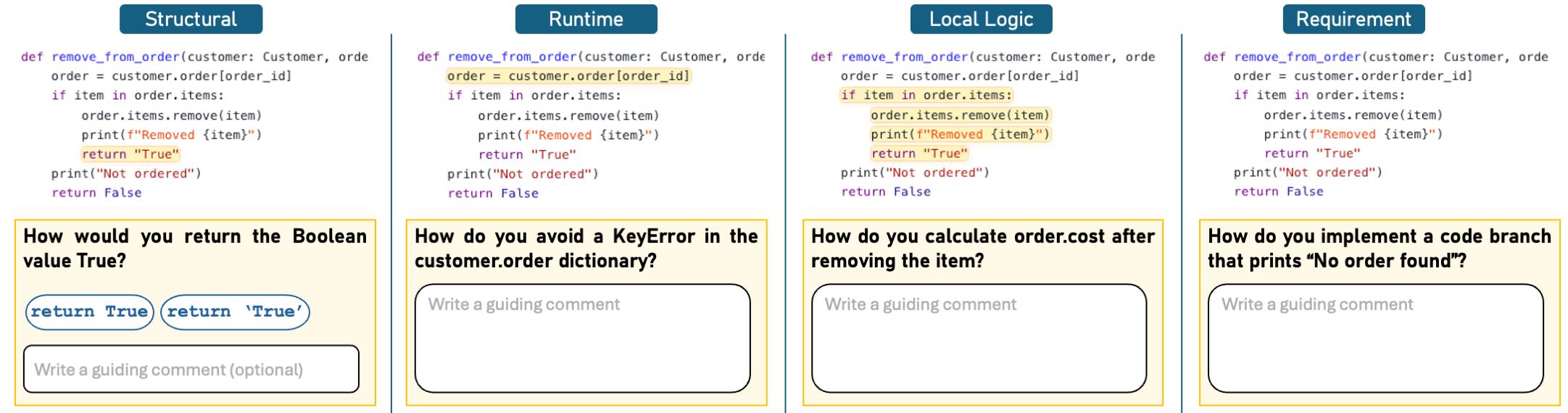}
    \caption{Example of a helper’s view displaying issue highlighting and scaffolding interactions across different levels of the Help Ladder for the function \texttt{remove\_from\_order}. Structural guidance focuses on surface-level syntax or representation issues. Runtime guidance addresses execution-time failures. Local Logic guidance targets behavioral correctness within the immediate implementation context. Requirement guidance addresses higher-level task expectations and missing behavioral branches. Since the requirement-level issue corresponds to missing functionality rather than an incorrect existing implementation, no specific code region is highlighted. The system supports lightweight multiple-choice interventions for quick corrections. For higher-context issues, helpers provide open-ended textual guidance that is inserted as contextual code comments on the helpee's editor. The question prompt is grounded in the focus of each level.}
    \label{fig:levels}
\end{figure*}

\begin{itemize}
    \item \textbf{Level 1: Structural Issues.} These include syntax errors and formatting mistakes. Such issues typically require minimal contextual understanding because they can be localized to specific lines of code.

    \item \textbf{Level 2: Runtime Issues.} These issues can cause failures during execution, such as null-pointer exceptions, invalid dictionary access, division-by-zero errors, or infinite loops. 

    \item \textbf{Level 3: Local Logic Issues.} These involve localized reasoning mistakes where the program executes successfully but produces incorrect outputs. Examples include off-by-one errors, incorrect variable resetting, or faulty conditional logic. Addressing these issues requires helpers to trace variable flow and understand the local implementation logic.

    \item \textbf{Level 4: Requirement Issues.} These represent the scenario where the implementation fails to satisfy the intended task goal or assignment requirements. Examples include selecting an inappropriate algorithm, failing to handle important edge cases, or misunderstanding the underlying problem specification. Resolving these issues requires helpers to reason about broader task intent and high-level program design.
\end{itemize}

The Help Ladder progressively surfaces programming obstacles based on these categories. Helpers are initially presented with lower-context entry points at lower level, enabling them to make quick contributions without needing to fully understand the teammate’s implementation. As helpers engage further, the system incrementally reveals deeper layers of contextual information and more abstract reasoning challenges.

\subsubsection{Expertise-Adaptive Helper Questioning}

To encourage targeted and actionable peer support, \sys{} frames helper prompts around focused ``\textit{How}'' questions that guide potential helpers toward implementation-oriented suggestions rather than generic advice. The system maintains a real-time model of each user’s programming experience by tracking concepts they have previously implemented in their personal editor. Based on this model, \sys{} adapts the granularity and structure of the questions and assistance shown to helpers (DG3).

For lower-context issues, particularly Level 1 and some Level 2 problems, helpers who have previously demonstrated familiarity with the relevant programming concepts may receive multiple-choice implementation suggestions (e.g., \texttt{print("testtest")}, \texttt{print("test test")}, or \texttt{print("test | test")}). In contrast, when helpers have not demonstrated prior understanding of the concept, \sys{} withholds AI-generated answer choices and instead provides open-ended response fields. This design discourages helpers from simply relaying AI-generated solutions they may not understand \cite{Margulieux02012016, dbox} while encouraging collaborative reasoning around the underlying programming concepts.

\subsubsection{Progressive Reveal}

To further reduce orientation cost, \sys{} visually grounds surfaced guidance within the teammate’s editor context (DG4). Suggestions are mapped directly to relevant code regions and functions through highlighting \cite{Ivie}, helping potential helpers quickly orient themselves within unfamiliar implementations while minimizing disruptive context switching. Rather than exposing all detected issues at once, \sys{} reveals suggestions progressively, beginning with lower-level issues that require less contextual understanding. When multiple issues appear at the same level, \sys{} prioritizes those most closely related to the console output. If multiple issues are equally relevant, the system orders them by their location in the code, from top to bottom.

\subsubsection{Sharing Suggestions with Helpees}

\begin{figure}
    \centering
    \includegraphics[width=1\linewidth]{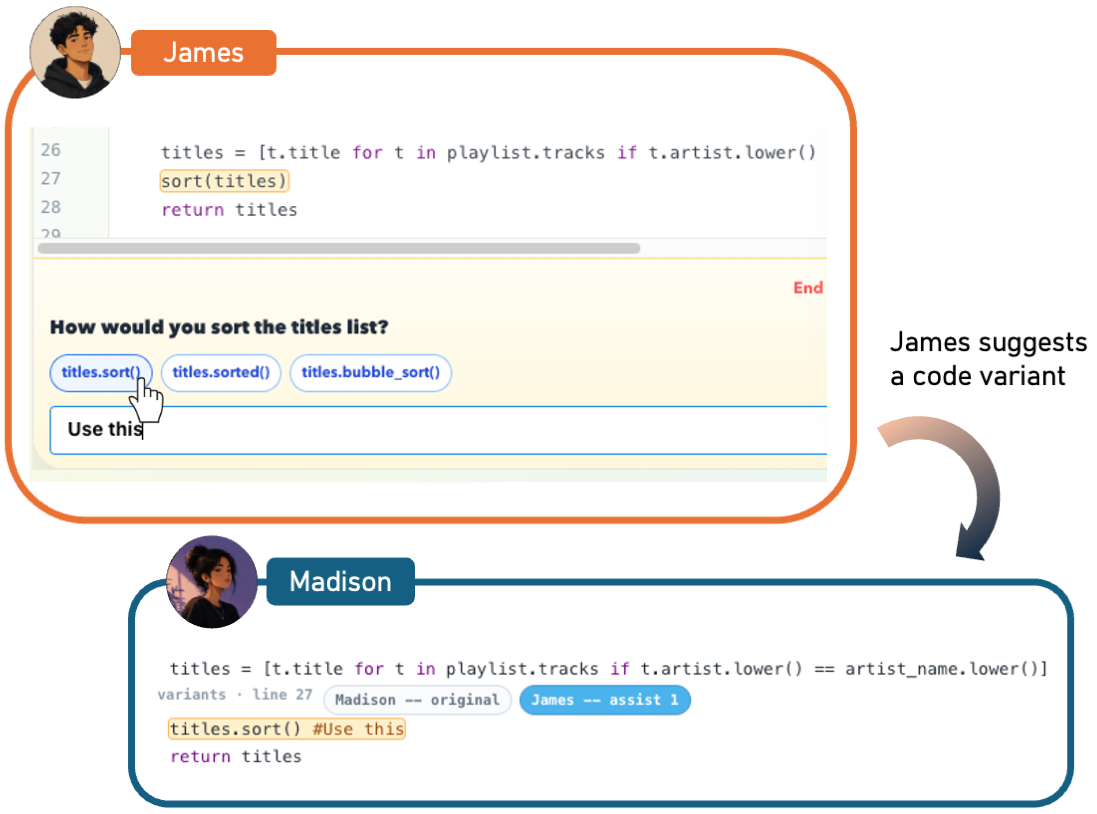}
    \caption{Example of low-context peer scaffolding through a multiple-choice suggestion. James identifies the appropriate method for sorting the \texttt{titles} list and submits the recommendation as a code variant to Madison’s code. James can also provide additional textual guidance, which is inserted as contextual comments in Madison’s personal editor.}
    \label{fig:codeVariantInteraction}
\end{figure}

As described previously, \sys{} supports two forms of peer scaffolding suggestions: multiple-choice code suggestions and open-ended guidance. Figure~\ref{fig:codeVariantInteraction} illustrates the process of sharing assistance between teammates.

For multiple-choice code suggestions, once a helper selects a suggestion, the chosen implementation is applied directly to the corresponding region of the helpee’s code as a new variant~\cite{Variolite}. \sys{} visually represents these alternative implementations inline within the editor, allowing helpees to switch between and compare different versions. When tests are executed, the currently displayed variant is used during execution. By introducing suggestions as lightweight code variants rather than directly overwriting existing implementations, \sys{} minimizes disruption to the helpee’s ongoing workflow while still enabling rapid peer contributions (DG2). Helpers may also attach short explanatory messages to provide additional rationale or clarification.

For open-ended guidance, helpers are provided with a free-form response interface. These responses are inserted as contextual comments adjacent to the associated code region or function, enabling helpers to share higher-level reasoning and implementation guidance without directly modifying the code.

\section{Evaluation}

Our evaluation focused on the effectiveness of the structured Help Ladder in supporting peer scaffolding.

\subsection{Methodology}

We conducted an in-lab within-subject study with 12 CS students (9 male, 3 female) to evaluate \sys{}, similar to prior work \cite{skWiki, collabIUI}. Participants were recruited through internal mailing lists and the institutional LMS with assistance from several course instructors. The participants were organized into four groups of three members each (e.g., G1-A to G1-C for Group 1). Additional participant background is summarized in Appendix~\ref{app:background}. The study employed a counterbalanced design to compare two conditions: (1) the full \sys{} system with the skill-adaptive help ladder enabled, and (2) a baseline condition without the adaptive peer scaffolding features. Apart from the Help Ladder, the two conditions shared the same user interface, help suggestion notification mechanism, and all other system functionalities, allowing us to isolate the contribution of the Help Ladder.

All participants provided informed consent prior to the session. Each study session was approximately 1.5 hours in duration. Sessions were audio-video recorded, and participants’ screens were captured for subsequent analysis. To establish rapport, we began each session with a brief ice-breaking activity, followed by a 5--10 minute technical refresher on Python programming. This refresher focused specifically on Python classes and object-oriented principles, as these were central to the study tasks. Participants were also provided with a Python cheatsheet\footnote{https://realpython.com/cheatsheets/python} for review.

Participants completed CP tasks under both system conditions. To mitigate learning effects, the order of system conditions and the specific programming tasks were counterbalanced across groups. For each condition, participants first received a brief introduction to the system’s interface and features, followed by a 5-minute warm-up task to ensure familiarity with the interface. This was then followed by a 20-minute collaborative session focused on the primary study task, during which data were collected. During the task, participants were not allowed to refer to external materials, including the cheatsheet or AI coding assistants.

Following each condition, participants completed a survey reflecting on their collaboration workload, adapted from~\cite{twlq}. Each session concluded with a brief semi-structured interview with individual participants to gather additional qualitative feedback on the system. The study protocol was approved by our institution’s IRB, and participants received \$25 compensation upon completing the study.

\subsection{Results}

We analyzed the recordings and study artifacts. The results are as follows.

\begin{table*}[t]
\centering
\caption{Comparison of collaboration behaviors between the baseline condition and \sys{} across groups}
\label{tab:collabResults}
\small
\begin{tabular}{lcccccccccc}
\toprule
& \multicolumn{2}{c}{Help Sessions}
& \multicolumn{2}{c}{Issues Identified}
& \multicolumn{2}{c}{Issues Resolved}
& \multicolumn{2}{c}{Time-to-Intervention (min)}
& \multicolumn{2}{c}{Tasks Completed} \\
\cmidrule(lr){2-3}
\cmidrule(lr){4-5}
\cmidrule(lr){6-7}
\cmidrule(lr){8-9}
\cmidrule(lr){10-11}
Group & Baseline & \sys{} & Baseline & \sys{} & Baseline & \sys{} & Baseline & \sys{} & Baseline & \sys{} \\
\midrule
G1 & 1 & 6 & 1 & 8 & 1 & 5 & 3.21 & 1.02 & 0 & 4 \\
G2 & 4 & 7 & 4 & 9 & 2 & 6 & 3.55 & 1.16 & 2 & 3 \\
G3 & 5 & 6 & 5 & 10 & 2 & 7 & 1.54 & 0.48 & 0 & 0 \\
G4 & 1 & 5 & 1 & 10 & 1 & 7 & 4.85 & 1.29 & 0 & 1 \\
\midrule
Mean & 2.75 & \textbf{6.00}* & 2.75 & \textbf{9.25}** & 1.50 & \textbf{6.25}** & 3.29 & \textbf{0.99}* & 0.50 & \textbf{2.00} \\
\bottomrule
\end{tabular}
\vspace{0.5em}
\begin{flushleft}
\footnotesize
Note: Asterisks indicate paired t-test comparisons between baseline and \sys{} across groups: * \(p < .05\), ** \(p < .01\).
\end{flushleft}
\end{table*}

\begin{table}[t]
\centering
\caption{Helping strategies relative to identified issues}
\label{tab:scaffoldingQuality}
\small
\setlength{\tabcolsep}{4pt}
\begin{tabular}{lcc}
\toprule
Helping Strategy & Baseline & \sys{} \\
\midrule
Direct-answer assistance & 5/11 (45.45\%) & 21/37 (56.76\%) \\
Hinting & 1/11 (9.09\%) & 12/37 (32.43\%) \\
Context gathering & 10/11 (90.91\%) & 9/37 (24.32\%) \\
Root-causing support & 6/11 (54.55\%) & 23/37 (62.16\%) \\
Explanatory scaffolding & 2/11 (18.18\%) & 10/37 (27.03\%) \\
\bottomrule
\end{tabular}
\end{table}

\subsubsection{Scaffolding Frequency and Quality}

As shown in Table~\ref{tab:collabResults}, groups using \sys{} participated in an average of 6.00 collaborative help sessions, compared to 2.75 sessions in the baseline. Across all four groups, the Help Ladder appeared to support more frequent transitions from independent work into collaborative problem solving. One exception was G3, which had a similar number of help sessions across the two conditions. This may be because participants in G3 already knew each other and were therefore more willing to reach out for help even without system support. Taken together, these observations are also consistent with our preliminary study, in which teams averaged only 1.25 help sessions and 0.50 resolved issues, suggesting significant improvements in collaborative support with the Help Ladder.

\sys{} also supported the identification of more issues compared to the baseline. In the baseline condition, the number of issues identified closely matched the number of help sessions, suggesting that each help interaction typically focused on a single visible problem. In contrast, \sys{} surfaced more issues than the number of help sessions because the Help Ladder decomposed programming obstacles into multiple layers. This decomposition enabled helpers to distribute their effort across more specific sub-issues. However, identifying issues did not always mean that participants were able to resolve them, as the number of resolved issues remained lower than the number of identified issues. Even so, groups using \sys{} resolved slightly more issues than the number of help sessions, whereas baseline groups resolved fewer issues than the number of help sessions. This suggests that, with \sys{}, a single help session could support more sustained engagement and lead to progress on multiple aspects of a helpee’s problem.

The average time-to-intervention, defined as the duration between a student becoming stuck and a teammate initiating help, also decreased from 3.29 minutes in the baseline condition to 0.99 minutes with \sys{}. This reduction may be explained by the lightweight entry points provided by \sys{}: more than half of the first suggested help opportunities were multiple-choice interventions, making it easier for helpers to respond quickly. Several participants described how the layered scaffolding reduced the uncertainty associated with helping. G1-C explained that ``\textit{the assistance feature made it easier to help my teammates because I didn't need to completely understand the issue at hand to help.}'' Similarly, G4-B reported that the system made it ``\textit{easy to identify when to start helping a teammate.}''

To further characterize the quality of peer scaffolding, we analyzed the helping strategies used during recorded help sessions. As shown in Table~\ref{tab:scaffoldingQuality}, we categorized helpers' strategies into \textit{direct-answer assistance} (explicitly providing the fix), \textit{hinting} (suggesting the direction of the fix), \textit{context gathering} (gathering information to understand the teammate's problem), \textit{root-causing support} (identifying the underlying cause), and \textit{explanatory scaffolding} (explaining relevant programming concepts). Overall, helpers in the \sys{} condition more frequently provided targeted assistance while requiring less context gathering than in the baseline. This trend was also observed in G3, where no tasks were ultimately completed under either condition: participants exhibited more explanatory scaffolding (5 vs. 1) while requiring less context gathering before providing assistance (2 vs. 5).

\subsubsection{Identification and Resolution of Programming Obstacles}

Participants identified and resolved more programming obstacles in the \sys{} condition than in the baseline condition. Further analysis of the code artifacts showed that the Help Ladder supported different categories of obstacles in distinct ways. Structural and execution errors were often addressed through low-context interactions with multiple-choice code variants, allowing helpers to provide quick corrections without needing extensive understanding of the surrounding implementation. These low-context interactions accounted for roughly two-thirds of all initial interventions, suggesting that participants often encountered small implementation barriers, such as forgetting specific Python operations or syntax patterns, that could be addressed through lightweight scaffolding. In the baseline condition, participants had to explicitly ask one another for help, which made collaboration more dependent on group dynamics. For example, G3 participants already knew each other and were comfortable reaching out, while G2 had a participant, G2-A, who acted as an icebreaker by initiating conversation and making the team more communicative. In contrast, G1 and G4 were quieter groups. Participants in these groups often continued working independently even when they were stuck, resulting in only one help interaction near the end of the baseline session. With \sys{}, however, the system reached out to potential helpers on behalf of helpees, which made help opportunities more visible and contributed to more issues being identified and resolved in G1 and G4.

After Level 1 issues were resolved, success rates for higher-level issues dropped substantially, with fewer than 20\% of these issues being resolved. As obstacles became more complex, participants often began investigating the problem but did not always complete the scaffolding process. One successful case occurred in G2 on the \texttt{view\_order\_summary} task, where the expected output required several non-obvious structural steps. G2-C was able to identify the missing line needed to match the expected output, demonstrating how the Help Ladder could support deeper reasoning when helpers were able to engage with the relevant context. However, lower engagement at higher levels may have also resulted from the absence of predefined choices, as helpers had to manually compose suggestions. As G3-C explained, ``\textit{I didn't know what to write...and I wasn't sure whether I was right or not.}''

\subsubsection{Collaboration Workload}

\begin{figure}
    \centering
    \includegraphics[width=1\linewidth]{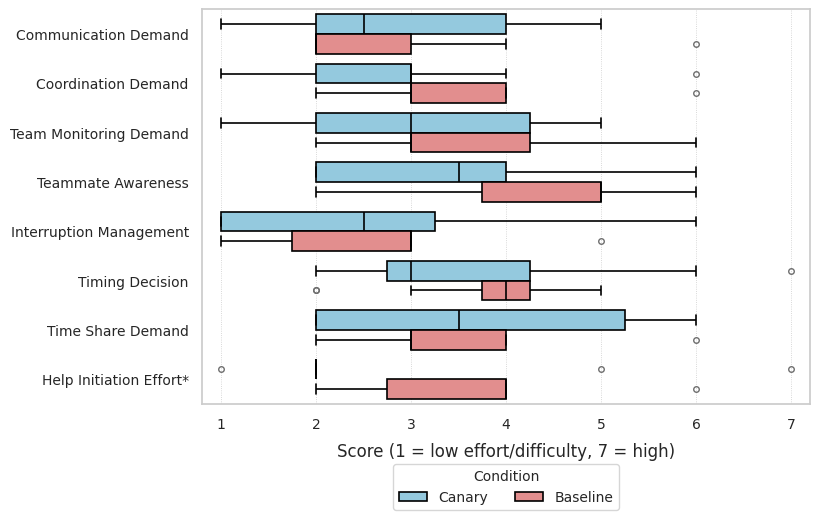}
    \caption{Comparison of collaboration-specific workload ratings~\cite{twlq} between the baseline condition and \sys{}. Lower scores indicate lower perceived effort or difficulty. Asterisks (*) indicate statistically significant differences under paired Wilcoxon signed-rank tests ($p < .05$).}
    \label{fig:surveyPlot}
\end{figure}

Survey results indicate that \sys{} reduced several collaboration-specific workload dimensions, particularly those related to recognizing and initiating help (Fig.~\ref{fig:surveyPlot}). Participants reported lower teammate awareness effort with \sys{} than in the baseline condition (mean 3.33 vs. 4.33, close to statistical significance of $p=.062$). Help initiation effort showed the strongest reduction and reached statistical significance (mean 2.58 vs. 3.58, $p=.016$). Coordination demand, team monitoring, and timing decision difficulty were also descriptively lower with \sys{} (2.92 vs. 3.42, 3.08 vs. 3.50, and 3.58 vs. 3.83, respectively), although these differences did not reach statistical significance. Together, these results suggest that \sys{} primarily reduced the effort associated with identifying collaboration opportunities and deciding when to intervene, rather than reducing all aspects of collaborative work equally. Qualitative feedback aligned closely with these patterns. Multiple participants described that the system reduced the effort required to determine when and how to intervene. G3-B stated that ``\textit{seeing the suggestions made it very intuitive who needed help and how to contribute}.'' Several participants also emphasized that the layered request structure reduced uncertainty around entering an ongoing task. Rather than interrupting a teammate and requiring a full explanation of the problem, participants could progressively inspect the issue through smaller contextualized steps. 

At the same time, not all collaboration workload dimensions decreased. Communication demand was identical across conditions (2.75 vs. 2.75), while interruption management and time-share demand were slightly higher with \sys{} (2.75 vs. 2.58 and 3.83 vs. 3.33). This pattern is consistent with the behavioral results: \sys{} increased the number of help sessions, which meant participants more frequently balanced their own coding work with active peer support. Rather than working independently for long uninterrupted periods, participants needed to continuously engage in collaboration, switch contexts, and divide attention between their own tasks and teammates' requests. As a result, some forms of coordination overhead naturally increased alongside the increase in collaborative engagement.

We also collected individual NASA-TLX workload measures to understand how participants perceived the overall task demands while using \sys{}. Results showed that participants perceived similar overall mental demand across conditions (3.58 vs. 3.17), although several dimensions trended slightly higher with \sys{}. In particular, temporal demand (3.67 vs. 2.83), effort (4.25 vs. 3.67), and frustration (3.83 vs. 2.67) were descriptively higher when using the system. Qualitative feedback suggests that this increase was closely tied to the increased collaborative activity enabled by \sys{}. Participants reported that the additional signals and collaboration opportunities required more continuous attention and monitoring compared to the baseline condition. Several participants explained that they often began helping with problems they initially understood well, but later encountered more complex issues that became difficult to resolve fully. This occasionally created frustration because participants felt uncertain about how long or how deeply they should continue assisting once they had joined the support process. As G4-A described, ``\textit{I don't know when to stop helping once it gets more complicated},'' reflecting the tension between supporting teammates and returning to their own tasks. At the same time, physical demand remained very low in both conditions (1.50 vs. 1.42), while perceived performance ratings were comparable overall, with participants reporting slightly better perceived performance when using \sys{} (4.50 vs. 5.00). These perceptions were also reflected in the behavioral outcomes. For example, in G2 under the baseline condition, G2-B was the only participant able to independently solve two tasks, whereas under \sys{}, all group members were able to successfully complete at least one task. This suggests that although participants experienced slightly higher coordination and attention demands with \sys{}, they also perceived stronger collective progress and broader task participation across the team.

\section{Discussion}

\subsection{Timely Actions in Real-Time Systems}

A common challenge in real-time systems is keeping up with continuously evolving information. In synchronous collaboration, delays in decision-making can lead to missed opportunities for coordination and support. Our findings suggest that prioritizing low-cost, immediate actions can facilitate more responsive collaboration. This insight extends beyond RCP to other real-time systems, where enabling fast, actionable entry points can improve interactions and better support users’ timely needs. In this context, AI can play a key role by identifying relevant information and presenting it in a timely and digestible manner to support rapid decision-making.

\subsection{Progressive Engagement}

Our work also demonstrates the value of progressive reveal in supporting RCP. By structuring assistance into layers of increasing complexity, users can gradually build understanding while remaining engaged in the task. This approach allows participants to make meaningful contributions early on and develop confidence before tackling more complex decisions. This suggests that progressive scaffolding can shape not only whether collaboration occurs, but also how collaborators contribute once an interaction begins. In time-sensitive settings, designing systems around incremental engagement may help users move more efficiently from initial participation toward deeper reasoning about complex information.

\subsection{Individual and Collaboration Workload}

Our findings highlight an inherent tension between progressing on one’s own task and actively collaborating with teammates. While \sys{} reduced several collaboration barriers and increased helping behaviors, collaboration itself can introduce additional workload for individuals. This pattern was reflected in the survey results, where participants reported slightly higher perceived workload in several NASA-TLX dimensions while using the system. In particular, participants needed to divide attention between their own programming progress and opportunities to support others. This suggests that collaborative support systems must carefully balance the benefits of peer assistance against the cognitive and temporal costs introduced by increased collaboration. If collaboration demands become too burdensome, participants may avoid engaging with teammates altogether. Conversely, if helping others is sufficiently lightweight and the perceived rewards are clear, users may be more willing to contribute actively to team success. In our study, the activity was framed as a shared team project in which participants collectively submitted a single assignment. As a result, participants could directly perceive the benefits of supporting one another, since helping teammates also contributed to the overall success of the group. This framing may have encouraged participants to tolerate the additional workload associated with collaboration in exchange for improved collective progress.

\section{Limitations \& Future Work}

This work has several limitations. First, our evaluation was conducted in a controlled, in-lab setting rather than in an authentic classroom environment. Our next step is to deploy \sys{} in live classroom settings to examine its impact on collaboration and potentially learning outcomes of that class.

Second, \sys{} has not yet been evaluated on more complex collaborative programming scenarios, such as large-scale codebases or projects requiring longer-term collaboration. While our evaluation focused on classroom programming challenges, future work should investigate the system's applicability to broader collaborative programming contexts.

Third, this study did not systematically control for individual dispositional differences or prior familiarity among participants. Groups were formed from the pool of students who signed up for the study, and some participants already knew each other beforehand (e.g., all participants in G3 were previously acquainted). Individual factors such as personality traits (e.g., extroversion and introversion), prior familiarity, and communication styles may have influenced students' willingness to initiate peer scaffolding and their overall interaction patterns. For example, although participants in G2 did not know each other prior to the study, G2-A frequently initiated verbal discussions that encouraged broader group collaboration. In contrast, participants in G1 rarely spoke verbally and instead relied more heavily on the text-comment scaffolding features to communicate and coordinate. These differences suggest that collaboration styles may vary substantially across teams depending on personality and social dynamics. Future work should investigate how individual differences and prior social relationships interact with collaborative intervention mechanisms in real-time programming environments.

\section{Conclusion}

In this paper, we examined challenges in real-time collaborative programming. Through formative studies, we found that students tend to persist in working individually even when they are stuck, hesitating to seek or offer help due to uncertainty about when and how to intervene under time constraints. We therefore introduced \sys{}, a collaborative IDE that supports peer scaffolding through LLM-powered layered help requests. By decomposing programming obstacles into progressively complex steps and surfacing localized entry points for potential helpers on behalf of helpees who are stuck, \sys{} lowers the barrier to participation and enables incremental engagement in the helping process. Our evaluation shows that this approach leads to more frequent and effective collaboration, improved issue resolution, and reduced perceived collaboration workload. By shifting AI assistance from solving problems directly to facilitating human-to-human support, \sys{} demonstrates a direction for designing AI-assisted collaborative tools that enhance, rather than replace, peer learning.

\section*{Acknowledgment}

We would like to thank Marcus Huynh for his assistance in conducting several user studies. We also thank Dr. David H. Smith IV, Wei-Lu Wang, Yuhang Zheng, Yi Fang, Hamid Tarashiyoun, and other colleagues for their feedback on this research.

\bibliographystyle{IEEEtran}
\bibliography{references}

\appendices

\section{Participant Background}
\label{app:background}

Table~\ref{tab:participantBackground} summarizes the background of the participants who took part in the evaluation. Participants were assigned to four groups of three members each.

\begin{center}
\captionof{table}{Participant background information for the evaluation study.}
\label{tab:participantBackground}
\small
\setlength{\tabcolsep}{3.2pt}
\begin{tabular}{cccccc}
\toprule
\textbf{Group} & 
\textbf{Participant} & 
\textbf{Gender} & 
\textbf{Age} & 
\makecell[c]{\textbf{Self-rated}\\\textbf{Python (1--7)}} &
\makecell[c]{\textbf{Prior}\\\textbf{Familiarity}} \\
\midrule
G1 & G1-A & Male & 21 & 4 & G1-C \\
   & G1-B & Male & 21 & 4 & -- \\
   & G1-C & Male & 20 & 5 & G1-A \\
\midrule
G2 & G2-A & Female & 20 & 4 & -- \\
   & G2-B & Male & 21 & 4 & -- \\
   & G2-C & Female & 21 & 4 & -- \\
\midrule
G3 & G3-A & Male & 25 & 5 & G3-B, G3-C \\
   & G3-B & Male & 24 & 5 & G3-A, G3-C \\
   & G3-C & Male & 30 & 3 & G3-A, G3-B \\
\midrule
G4 & G4-A & Male & 21 & 5 & -- \\
   & G4-B & Female & 21 & 4 & -- \\
   & G4-C & Male & 21 & 4 & -- \\
\bottomrule
\end{tabular}
\end{center}

\end{document}